\documentclass[draft]{agujournal}
\journalname{Geophysical Research Letters}
\usepackage{lineno}

\begin{document}

%% ------------------------------------------------------------------------ %%
%
%  TITLE
%
%% ------------------------------------------------------------------------ %%

% 10 figures, 6500 words limit, 50 references
\title{Axial Seamount: Periodic tidal loading reveals stress dependence of the earthquake size distribution (\textit{b} value)}

%% ------------------------------------------------------------------------ %%
%
%  AUTHORS AND AFFILIATIONS
%
%% ------------------------------------------------------------------------ %%

 \authors{Yen Joe Tan\affil{1},
 Felix Waldhauser\affil{1}, Maya Tolstoy\affil{1}, and William S. D. Wilcock\affil{2}}
 
\affiliation{1}{Lamont-Doherty Earth Observatory of Columbia University, Palisades, New York 10964, USA.}
\affiliation{2}{School of Oceanography, University of Washington, Seattle, WA
98195, USA.}

%% Corresponding Author
%(include name and email addresses of the corresponding author.  More
%than one corresponding author is allowed in this Word file and for
%publication; but only one corresponding author is allowed in our
%editorial system.)  

\correspondingauthor{Yen Joe Tan}{yjt@ldeo.columbia.edu}

%  List up to three key points (at least one is required)
%  Key Points summarize the main points and conclusions of the article
%  Each must be 100 characters or less with no special characters or punctuation 

\begin{keypoints}
\item Above a threshold stress amplitude, \textit{b} value decreases with increasing tidal stress.
\item \textit{b} value varies by $\sim$0.09 per kPa change in Coulomb stress.
\item \textit{b} values can be used to map small stress variations in the Earth's crust.
\end{keypoints}

%% ------------------------------------------------------------------------ %%
%
%  ABSTRACT
%
%% ------------------------------------------------------------------------ %%

\begin{abstract}
Earthquake size-frequency distributions commonly follow a power law, with the \textit{b} value often used to quantify the relative proportion of small and large events. Laboratory experiments have found that the \textit{b} value of microfractures decreases with increasing stress. Studies have inferred that this relationship also holds for earthquakes based on observations of earthquake \textit{b} values varying systematically with faulting style, depth, and for subduction zone earthquakes, plate age. However, these studies are limited by small sample sizes despite aggregating events over large regions, which precludes the ability to control for other variables that might also affect earthquake \textit{b} values such as rock heterogeneity and fault roughness. Our natural experiment in a unique seafloor laboratory on Axial Seamount involves analyzing the size-frequency distribution of $\sim$60,000 microearthquakes which delineate a ring-fault system in a 25 km$^3$ block of crust that experiences periodic tidal loading of $\pm$20 kPa. We find that above a threshold stress amplitude, \textit{b} value is inversely correlated with tidal stress. The earthquake \textit{b} value varies by $\sim$0.09 per kPa change in Coulomb stress. Our results support the potential use of \textit{b} values to estimate small stress variations in the Earth's crust. 
\end{abstract}

%% ------------------------------------------------------------------------ %%
%
%  TEXT
%
%% ------------------------------------------------------------------------ %%

%%% Suggested section heads:
% \section{Introduction}
% 
% The main text should start with an introduction. Except for short
% manuscripts (such as comments and replies), the text should be divided
% into sections, each with its own heading. 

% Headings should be sentence fragments and do not begin with a
% lowercase letter or number. Examples of good headings are:

% \section{Materials and Methods}
% Here is text on Materials and Methods.

% \subsection{A descriptive heading about methods}
% More about Methods.
% 
% \section{Data} (Or section title might be a descriptive heading about data)
% 
% \section{Results} (Or section title might be a descriptive heading about the
% results)
% 
% \section{Conclusions}

\section{Introduction}
\label{sec:intro}
Earthquake occurrence is primarily controlled by the stress state on fault interfaces. Because \textit{in situ} stress measurements are difficult to obtain, a proxy for estimating the stress state of fault zones through their seismic cycles is valuable for understanding earthquake occurrence and forecasting earthquakes. Earthquakes follow a power-law size-frequency distribution given as log$_{10}(N) = a - bM$, where \textit{N} is the number of earthquakes greater than or equal to magnitude \textit{M}, and \textit{a} and \textit{b} are constants \citep{Gut44}. The value \textit{a} describes the total number of earthquakes while the \textit{b} value describes the relative frequency of small and large magnitude earthquakes. In rock fracture experiments, acoustic emissions from small cracking events follow the same power-law size distribution \citep{Sch68}. Furthermore, their \textit{b} values have been found to decrease (larger proportion of large events) with increasing differential stress \citep{Sch68, Amit03, Goe13}. 

The same stress dependence of \textit{b} value has been inferred to apply to earthquakes. The \textit{b} value of earthquakes has been found to vary systematically with faulting style \citep{Sch05a}, depth \citep{Spa13}, and for subduction zone earthquakes, plate age \citep{Nish14}. These observations are consistent with the earthquake \textit{b} value decreasing with increasing differential stress \citep{Sch15}. However, these studies were restricted to using minimum bins of as few as 50 to 200 earthquakes to calculate the \textit{b} values, which is barely a large enough sample size to even establish the existence of a power-law distribution \citep{Stu12}. In addition, these studies had to aggregate events over large regions and thus were unable to control for other variables that might also affect earthquake \textit{b} values such as rock heterogeneity \citep{Mor97} and fault roughness \citep{Goe17}. Establishing whether earthquake \textit{b} value varies systematically with stress is critical for demonstrating its potential use as a stress meter in the Earth's crust which could help improve forecasting of large earthquakes \citep{Sch05b, Nan12, Gul16} and volcanic eruptions \citep{Kat15}.

Tidal forcing on the Earth produces periodic stress changes on the order of several kPa. Studies to establish a correlation between global earthquake rate and tidal stress changes have produced equivocal results \citep[e.g.][and references therein]{Em97} with mainly negative results in continental regions \citep{Vid98, Wang15}. However, \cite{Coch04} found statistically-significant tidal triggering for shallow, subduction-zone thrust earthquakes where stress changes due to ocean tidal loading can be an order of magnitude larger than the solid earth tides. Even stronger tidal triggering of earthquakes \citep{Wil01,Tol02,Str07} that weakens post-eruption \citep{Wil16,Tan18} has been documented at mid-ocean ridges. For instance, at Axial Seamount which is located at the intersection of the Juan de Fuca Ridge and the Cobb-Eickelberg hotspot, earthquakes occur preferentially during low ocean height \citep{Tol02,Wil16}. \cite{Sch18} recently demonstrated that the exponential increase in seismicity rate with tidal stress at Axial Seamount agrees with predictions of both rate-state and stress corrosion theories \citep[Fig. 5]{Sch18}, and the long-documented high sensitivity can be explained by the shallow depths of the earthquakes. 

As part of the Ocean Observatory Initiative (OOI), a cabled seismic network was installed on the summit of Axial Seamount (Fig. 1a) with time-corrected seismic data streaming from late January 2015 \citep{Wil16}. In the three months before the volcano erupted in April 2015, $\sim$60,000 earthquakes were located using the double-difference method \citep{Wald00}. The earthquakes delineate an outward-dipping ring-fault system that extends to $\sim$2-km depth \citep{Wil16} (Fig. 1). The large number of events located within a small region, combined with the earthquakes' sensitivity to tidal stress perturbations \citep{Wil16,Sch18}, make this an excellent natural laboratory to study how the earthquake \textit{b} value relates to stress changes.

\begin{figure}[ht!]
\centerline{\includegraphics[height=3.8in]{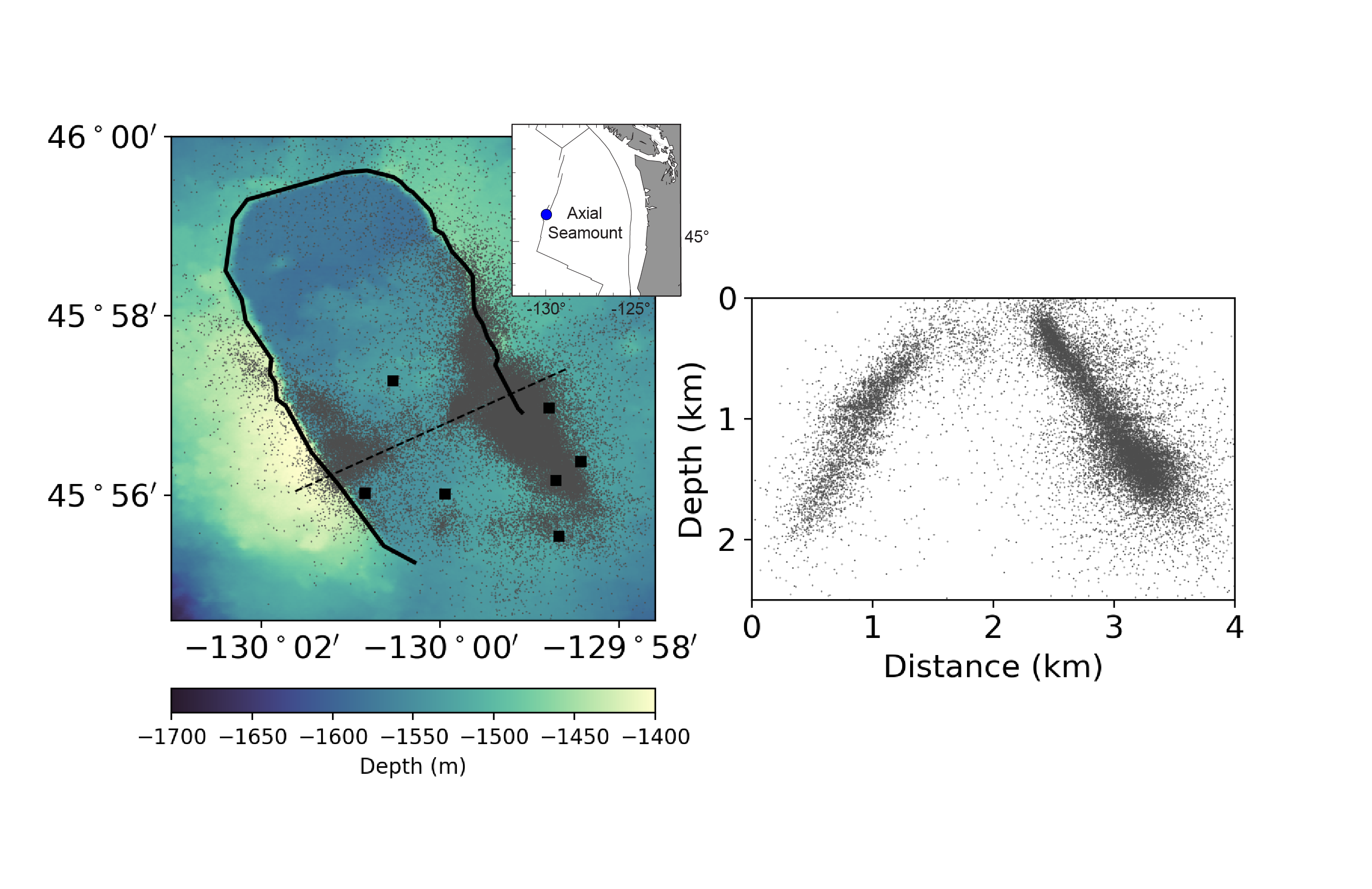}}
\caption{Locations of $\sim$35,000 earthquakes above $M_c$ = 0.1 between January 22nd and April 23rd 2015. \textbf{a}, Bathymetric map with earthquake epicenters (grey dots), seismometers (black filled squares), caldera rim (black line), and the cross section shown in (b) (black dashed line). Inset shows regional location of Axial Seamount. \textbf{b}, Depth cross-section across the caldera showing the projected earthquake locations within 0.5 km of the profile.}
\end{figure}

\section{Methods}
\label{sec:meth}
\subsection{Earthquake Catalog}
In the first year of operation, $\sim$70,000 earthquakes were located by the OOI Axial seismic network \citep{Wil16}. In the three months before the volcano erupted, the majority of the composite focal mechanisms determined showed normal or oblique-normal sense of motion. During the one-month-long eruption period, the slip direction was reversed as the volcano deflated \citep{Lev18}. After the eruption, the seismicity rate decreased substantially \citep{Wil16,Wil18} with the focal mechanisms suggesting heterogeneous fault slip directions \citep{Lev18}. Therefore, in this paper, we only examine the $\sim$60,000 earthquakes that occurred in the three months before the volcano erupted. The earthquake catalog, including the moment magnitudes ($M_W$) estimated following \citet{Tre83}, has been previously published \citep{Wil16}.

\subsection{\textit{b} Value}
We estimate the \textit{b} values using the maximum likelihood method \citep{Aki65}, accounting for the use of binned magnitudes \citep{Ut66}:
\begin{equation}
b = \frac{log_{10} e}{\bar{M} - \Big(M_{c}-\frac{\Delta M}{2}\Big)} ,
\end{equation}
where $M_{c}$ is the magnitude of completeness of the data set, $\bar{M}$ is the mean magnitude of earthquakes with magnitude $\geq M_{c}$, and $\Delta M$ is the binning interval of the magnitude, which is 0.1 in this study. We estimate the standard deviation of the \textit{b} value estimate following \citet{Shi82}:
\begin{equation}
\delta b = 2.3 b^{2} \sqrt{\frac{\sum \limits_{i}^n (M_{i} - \bar{M})^{2}}{n(n-1)}} ,
\end{equation}
where \textit{n} is the sample size. We quantify the significance of the \textit{b} value difference between two groups of earthquakes using Utsu's test \citep{Ut92}:
\begin{equation}
p \approx \exp \Bigg( \frac{-\Delta AIC}{2} - 2 \Bigg) ,
\end{equation}
\begin{equation}
\Delta AIC = -2 (N_{1} + N_{2}) \ln (N_{1} + N_{2}) + 2 N_{1} \ln \Bigg( N_{1} + \frac{N_{2}b_{1}}{b_{2}} \Bigg) + 2 N_{2} \ln \Bigg( N_{2} + \frac{N_{1}b_{2}}{b_{1}} \Bigg) - 2 ,
\end{equation}
where \textit{p} is the probability that the two groups of earthquakes are drawn from the same population, AIC is the Akaike's information criterion, $N_1$ and $N_2$ are the number of earthquakes, and $b_1$ and $b_2$ are the estimated \textit{b} values of the two groups of earthquakes. We also compare the \textit{b} value difference between two groups with their standard deviations \citep{Kag97}:
\begin{equation}
z = \frac{b_1 - b_2}{\sqrt{\sigma_1^2 + \sigma_2^2}},
\end{equation}
The null hypothesis that two \textit{b} values come from the same population can be rejected at the 95\% confidence level if \textit{z} exceeds 1.96 and at the 99\% confidence level if \textit{z} exceeds 2.58.

\subsection{Magnitude of Completeness}
We first estimate the magnitude of completeness ($M_c$) of the catalog using the point of maximum curvature of the frequency-magnitude distribution (FMD) \citep{Wie99}, which is equivalent to finding the magnitude bin with the highest number of earthquakes in the non-cumulative FMD \citep{Mig12}. We find $M_c$ = 0.0 (Fig. S1). We then estimate $M_c$ using the goodness-of-fit (GFT) method by comparing observed and synthetic cumulative FMDs \citep{Wie00}. We calculate synthetic cumulative FMDs using estimated \textit{a} and \textit{b} values of the observed earthquake catalog assuming a range of increasing cutoff magnitudes $M_{co}$. The goodness-of-fit is quantified using the parameter \textit{R}:
\begin{equation}
R_{M_{co}} = 100 - \Bigg(100 \frac{\sum_{M_{co}}^{M_{max}} |O_{i} - S_{i}|}{\sum_{M_{co}}^{M_{max}} O_{i}} \Bigg) ,
\end{equation}
where $O_{i}$ and $S_{i}$ are the observed and predicted number of earthquakes in each magnitude bin. $M_{c}$ is then the first $M_{co}$ where \textit{R} exceeds a fixed threshold, typically defined at $90\%$ level of fit \citep{Wie00} because real catalogs rarely achieve $95\%$ level of fit \citep{Woes05}. 
We obtain $M_c$ = $-0.1$ when using a $90\%$ fit threshold and $M_c$ = 0.1 when using a $95\%$ fit threshold (Fig. S1). Finally, we estimate $M_c$ base on the $b$ value stability as a function of assumed cutoff magnitude $M_{co}$ \citep{Cao02}. $M_{c}$ is the first $M_{co}$ at which $|b_{ave}-b| \leq \delta b$ \citep{Woes05}, with $b_{ave}$ being the mean of the $b$ values estimated for three successive $M_{co}$ (magnitude range of 0.3 since the bin interval is 0.1) and $\delta b$ being the standard deviation of the \textit{b} value estimate \citep{Shi82}. We obtain $M_c$ = 0.3 (Fig. S1). 

The maximum curvature (MAXC) and the GFT-90\% methods can underestimate $M_c$ \citep{Woes05} while the method based on $b$ value stability (MBS) may overestimate $M_c$ \citep{Mig12}. Therefore, in this study, we adopt $M_c$ = 0.1 from the GFT-95\% method which leaves us with $\sim$35,000 earthquakes above $M_c$ and an estimated \textit{b} value of 1.31 $\pm$ 0.01. We also consider the more conservative estimate of $M_c$ = 0.3 from the MBS method, which leaves us with $\sim$20,000 earthquakes above $M_c$ and an estimated \textit{b} value of 1.39 $\pm$ 0.01. The estimated \textit{b} values are consistent with previous observations of \textit{b} $>$ 1 for normal fault events \citep{Sch05a} and in marine volcanic environments \citep{Boh08}.

\subsection{Tidal Stress}
We estimate the horizontal strains due to body tides using the SPOTL software which assumes an elastic and spherical Earth (degree-two Love numbers $h$ = 0.6114, $k$ = 0.3040, and $l$ = 0.0832) and computes the tidal strains directly from the positions of the Moon and the Sun \citep{Ag97}. We then calculate the vertical strain from the horizontal strains assuming a plane stress condition:
\begin{equation}
\Delta \epsilon_{zz} = \frac{-\nu}{1-\nu} (\Delta \epsilon_{xx} + \Delta \epsilon_{yy}), 
\end{equation}
using Poisson's ratio $\nu$ of 0.23 which is consistent with $V_P = 5.4$ km/s, $V_S = 3.2$ km/s, and a density of 2800 kg/m$^3$. The Poisson's ratio quantifies the effect where a material tends to contract along the axes perpendicular to the axis of tensile strain. For the effects of ocean tidal loading, we first obtain the predicted tidal height for the eight major short-period tidal constituents (K1, K2, M2, N2, O1, P1, Q1, and S2) using the EOT11a global ocean tidal model \citep{Sav12} combined with the Oregon State University regional ocean tidal model for the west coast of the United States \citep{Eg02} as provided with the SPOTL software. We then calculate the ocean tidal loading effect in two parts. First, we estimate the horizontal strains due to variable regional ocean tidal loading using the SPOTL software, which uses a mass-loading Green's function for strain based on the Gutenberg-Bullen Earth model \citep{Ag97}. We then calculate the vertical strain from the horizontal strains assuming a plane stress condition before converting strains to stresses using elastic constants consistent with a Poisson's ratio of 0.23. Secondly, we estimate the vertical stress perturbation due to direct ocean tidal loading as
\begin{equation}
\Delta \sigma_{zz} = \rho g h, 
\end{equation}
where $\rho$ is the density of seawater (1030 kg/m$^3$), $g$ is the gravitational acceleration (9.8 m/s$^2$), and $h$ is the tidal height relative to its mean value. We then estimate the horizontal stresses from the vertical stress assuming uniaxial strain:
\begin{equation}
\Delta \sigma_{xx} = \Delta \sigma_{yy} = \frac{\nu}{1-\nu} \Delta \sigma_{zz}, 
\end{equation}
Finally, we combine the various tidal stress components to form the stress tensor. We find that at Axial Seamount, ocean tides are much larger than body tides and hence the vertical tidal stress dominates (Fig. S2). We calculate the tidal-stress time series in 5-minute intervals. We assume the stresses estimated at the seafloor applies to the earthquake source region because the tidal wavelengths are very long compared to the earthquake depths of mostly less than 2 km (Fig. 1b). We do not account for the effect of bathymetry in our tidal stress calculations. The bathymetry only varies by less than 200 m around our earthquake epicenter region (Fig. 1a) and while it might affect the absolute value of our tidal stress estimates, the relative difference in tidal stress at different times is expected to still be valid. 

We assume the earthquakes are predominantly normal faulting events since the ring-fault system appears to have accommodated pre-eruptive inflation \citep{Wil16}, 79\% (31 of 39) of the composite focal mechanisms determined before the eruption based on first-motion polarity showed normal or oblique-normal sense of motion \citep{Lev18}, and this is a region of tectonic extension. This initially seems to contradict the well-documented preferential occurrence of earthquakes during low ocean height at Axial Seamount \citep{Tol02, Wil16} because a decrease in ocean height (increase in tensile vertical stress) should produce a Coulomb stress change that inhibits slip on normal faults. However, \citet{Sch18} resolved this apparent paradox by accounting for the effect of the underlying magma chamber on the stress distribution. The higher compressibility of the magma chamber means that it will inflate or deflate relative to the surrounding crust in response to tidal stresses and produce Coulomb stress changes on the fault that is opposite in sign as those produced directly by the tidal stresses \citep[Fig. 3]{Sch18}. When the magma chamber bulk modulus is below a critical value, the magma chamber effect will exceed that of the direct tidal stress effect and the phase of the tidal triggering gets inverted \citep[Fig. 4]{Sch18}, as observed at Axial Seamount. Since the vertical tidal stress dominates at Axial Seamount (Fig. S2) and the average Coulomb stress change on the fault can be approximated as $\Delta CFS = \chi \sigma_{zz}$ with $\chi$ dependent on the magma chamber bulk modulus \citep{Sch18}, we will focus on variations in vertical stress due to the combined effects of ocean tidal loading and body tide. We adopt tension as positive and an increase in vertical stress represents an increase in encouraging stress (Coulomb stress change that favors slip on normal fault). The vertical tidal stress has estimated amplitudes of $\pm$20 kPa (Fig. S2).

\section{Results}
\label{sec:res}
We assign each earthquake a tidal stress value based on its origin time. After sorting the earthquakes based on their associated tidal stress values, we calculate the \textit{b} values for non-overlapping bins of 2,000 events. When using $M_c$ = 0.1, we also calculate the \textit{b} values for moving bins of 10,000 events, shifted by 5,000 events. When using $M_c$ = 0.3, we calculate the \textit{b} values for moving bins of 5,000 events, shifted by 2,500 events. For each bin, we re-estimate $M_c$ using the GFT-95\% method and only keep the data point if the re-estimated $M_c$ equals the $M_c$ of the bulk data. Only 3 out of 40 data points did not fulfill the criteria. We find that the earthquake \textit{b} value only decreases systematically with increasing tidal stress when stress amplitudes exceed a certain threshold (Fig. 2). 

\begin{figure}[ht!]
\centerline{\includegraphics[height=3.0in]{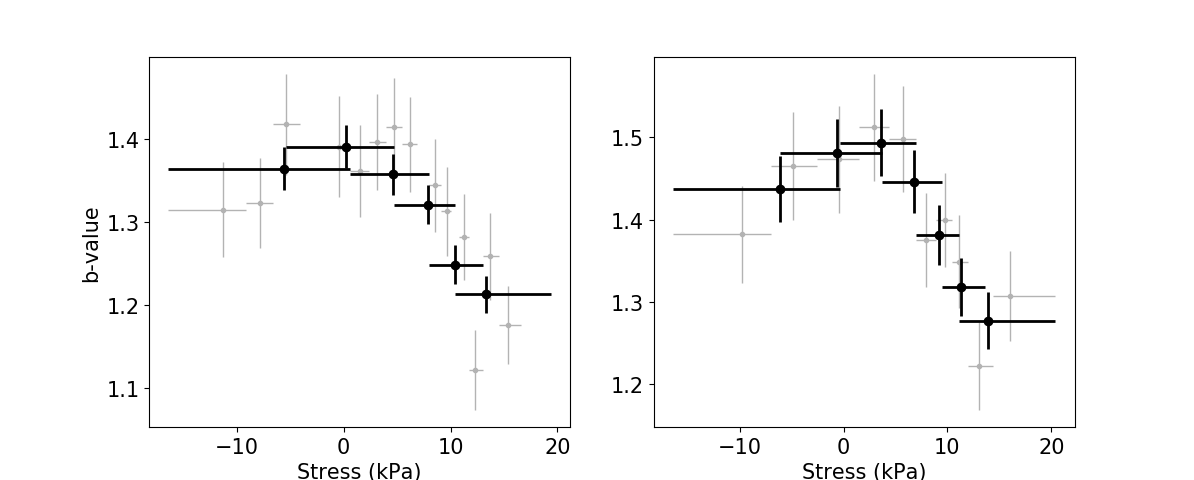}}
\caption{The earthquake \textit{b} value as a function of tidal stress. The vertical error bars represent two standard deviations of the estimated \textit{b} values \citep{Shi82}. The horizontal bars represent the range of earthquake tidal stress values included in each bin, with the markers centered at the mean earthquake tidal stress for each bin. \textbf{a}, Using $M_c$ = 0.1. Non-overlapping bins of 2,000 events (gray) as well as moving bins of 10,000 events shifted by 5,000 events (black). \textbf{b}, Using $M_c$ = 0.3. Non-overlapping bins of 2,000 events (gray) as well as moving bins of 5,000 events shifted by 2,500 events (black).}
\end{figure}

We further investigate the relationship between $b$ value and tidal stress by looking at how the $b$ value varies between fixed stress bins. We calculate the \textit{b} values for non-overlapping bins of 2 kPa for stress values between -10 and 16 kPa. We pick this range because it incorporates $\sim$92\% of the earthquakes (Fig. S3) and allows us to use a reasonably large number of events per stress bin. The number of events vary between stress bins (Fig. S3) so we adopt the following strategy to maintain consistency: For each stress bin, we estimate 1,000 \textit{b} values using events randomly drawn with replacement from the earthquake population. When using $M_c$ = 0.1, we draw 1,300 events for each \textit{b} value calculation because the stress bin with the smallest number of events contain $\sim$1,300 earthquakes (Fig. S3). When using $M_c$ = 0.3, we draw 700 events for each \textit{b} value calculation. The reported \textit{b} value is then the average \textit{b} value from the bootstrapping. We find that at low stress values, the $b$ values are high but remain relatively constant. However, at stress amplitudes greater than 5 kPa, the earthquake \textit{b} value decreases linearly with increasing tidal stress at $\sim$0.03 per kPa (Fig. 3). 

\begin{figure}[ht!]
\centerline{\includegraphics[height=3.0in]{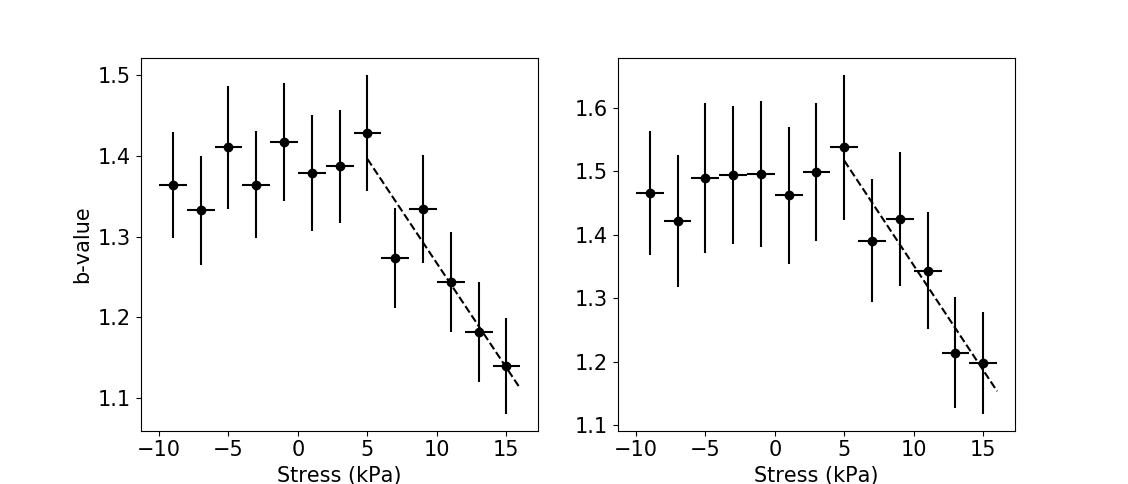}}
\caption{The earthquake \textit{b} values for non-overlapping stress bins of 2 kPa. The vertical error bars represent two standard deviations of the estimated \textit{b} values from bootstrapping. The horizontal bars represent the tidal stress range for each bin. Dashed lines represent linear least-squares fits, both giving \text{b} value varying by $\sim$ 0.03 per kPa. \textbf{a}, Using $M_c$ = 0.1. \textbf{b}, Using $M_c$ = 0.3.}
\end{figure}

We test the statistical significance of the \textit{b} value variations as follow: Using $M_c$ = 0.3, we sort the earthquakes based on their associated tidal stress values before splitting them into two equal-size groups. The lower tidal stress group has a mean stress of $-$1 kPa while the higher tidal stress group has a mean stress of 11 kPa. We further verified that the lower and higher tidal stress groups both have $M_c$ = 0.3. We then plot the cumulative and non-cumulative FMDs. The cumulative FMD curves show increasing separation at larger magnitudes. The non-cumulative FMD curves intersect at around $M_w$ = 0.5 (Fig. 4). These results show that the slopes of the FMD curves for the lower tidal stress group is steeper (larger \textit{b} value) than that of the higher tidal stress group. The lower tidal stress group has a \textit{b} value of 1.46 $\pm$ 0.01 while the higher tidal stress group has a \textit{b} value of 1.33 $\pm$ 0.01. We obtain similar \textit{b} values and standard deviations from bootstrapping. The \textit{b} value difference between the two groups are statistically significant at a $<$1\% level based on both the Utsu's test \citep{Ut92} and the \textit{z}-test (see Methods). 

The FMD curves deviate from linearity at large magnitudes. This could reflect a real departure from the power law distribution at large magnitudes or simply statistical fluctuations due to under-sampling. To quantify how this affects our results, we repeat the \textit{b} value calculations after excluding earthquakes of $M_w$ greater than 1.5 where the FMD curves become nonlinear \citep{Kag97} (Fig. 4). The lower tidal stress group now has an estimated \textit{b} value of 1.53 $\pm$ 0.01 while the higher tidal stress group has an estimated \textit{b} value of 1.40 $\pm$ 0.01. The \textit{b} value difference between the two groups remains statistically significant at a $<$1\% level based on both the Utsu's test \citep{Ut92} and the \textit{z}-test. We also repeat the analysis shown in Figure 3 and while the absolute \textit{b} values become larger, we obtain similar trends with the earthquake \textit{b} value decreasing linearly with increasing tidal stress at $\sim$0.03 per kPa when stress amplitude exceeds 5 kPa (Fig. S4). This is unsurprising because the maximum-likelihood estimate of the \textit{b} value uses the average earthquake magnitude (Eq. 1) and is therefore only slightly affected by the small number of large magnitude events. 

\begin{figure}[ht!]
\centerline{\includegraphics[height=3.0in]{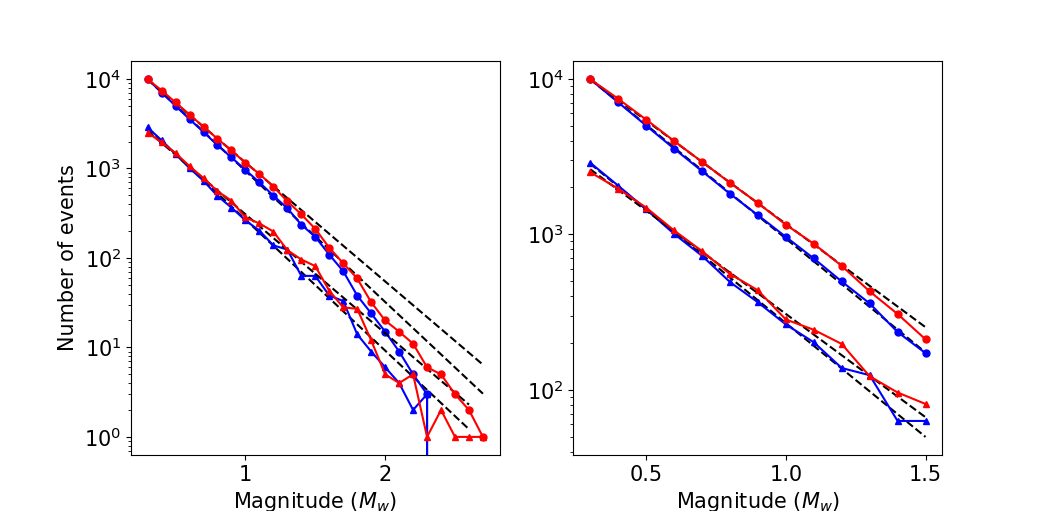}}
\caption{Cumulative (circle) and non-cumulative (triangle) FMDs of two groups of earthquakes. Dashed lines indicate maximum-likelihood fits to the data. \textbf{a}, Using $M_c$ = 0.3, the $\sim$20,000 earthquakes are split into two equal-size groups after being sorted based on their tidal stress values. The lower tidal stress group (blue) has a mean stress of $-$1 kPa while the higher tidal stress group (red) has a mean stress of 11 kPa. \textbf{b}, Zoom-in of a.}
\end{figure}

We further verified that the minimum bin sizes often used to document \textit{b} value variations \citep{Sch05a,Spa13,Nish14} are insufficient to robustly constrain our observed effect. We determine the minimum bin size needed to resolve the \textit{b} value variations we observe as follow: Using $M_c$ = 0.3, we sort the earthquakes based on their associated tidal stress values and split them into two equal-size groups (see Fig. 4). For a range of bin sizes, we then calculate 1,000 \textit{b} values using events randomly drawn without replacement from the original population. The reported \textit{b} value is then the average \textit{b} value from the bootstrapping with the associated uncertainties (Fig. 5). For the \textit{b} value difference between the lower and higher tidal stress groups to be statistically significant at a $<$5\% and $<$1\% level based on the Utsu's test \citep{Ut92}, we need minimum bin sizes of 900 and 1,600 respectively. For the \textit{b} value difference between the two groups to be statistically significant at a $<$5\% and $<$1\% level based on the \textit{z}-test, we need minimum bin sizes of 800 and 1,300 respectively. 

\begin{figure}[ht!]
\centerline{\includegraphics[height=4.0in]{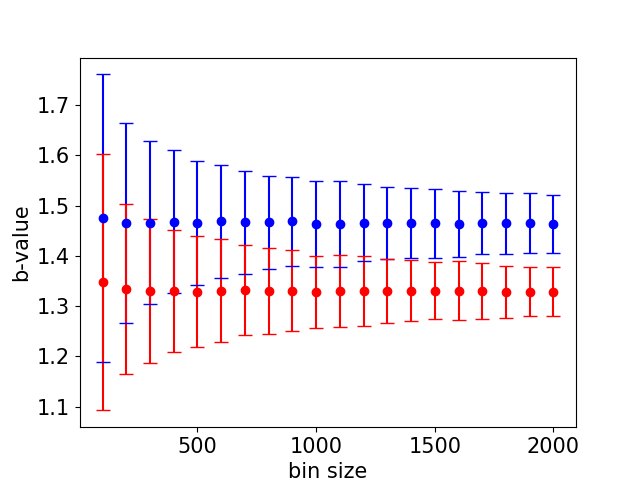}}
\caption{The estimated earthquake \textit{b} values for the lower (blue) and higher (red) tidal stress groups as a function of bin sizes. For each bin size, we calculate 1,000 \textit{b} values using events randomly drawn without replacement from the original population. The reported \textit{b} value is then the average \textit{b} value from the bootstrapping. The vertical error bars represent two standard deviations.}
\end{figure}

In multiple continental regions, earthquake \textit{b} values have been found to decrease with increasing depth which has been interpreted as the result of increasing crustal strength \citep{Spa13} and material homogeneity \citep{Mor97} with depth. We similarly find that at Axial Seamount, the earthquake \textit{b} value decreases with increasing depth (Fig. S5). Therefore, our observation of \textit{b} value decreasing with increasing tidal stress could simply reflect the average earthquake depth increasing with tidal stress. While the Chi-squared test suggests the earthquake depth distributions of the lower and higher tidal stress groups come from two different populations, the depth difference is such that the mean and median depths of the higher tidal stress group is shallower by $\sim$18 m and $\sim$16 m respectively. We also find that the mean and median earthquake depth decrease with increasing tidal stress (Fig. S6). Since \textit{b} value decreases with increasing depth, this change in depth distribution would have resulted in \textit{b} value increasing with tidal stress. Therefore, our observation of earthquake \textit{b} value decreasing with increasing tidal stress is unlikely to be a secondary effect of change in earthquake depth distribution with tides. 

The earthquake spatial distribution of the lower and higher tidal stress groups also differs slightly, with relatively more events on the western and northeastern walls of the caldera for the lower tidal stress group (Fig. S7). We verify that the \textit{b} value variation with tidal stress that we observe is still valid at smaller spatial scale as follow: We first bin the earthquakes into 1 km$^2$ spatial grids and find that there are five bins with more than 1,000 events (Fig. S7c) . For each of these five bins, we sort the earthquakes base on their tidal stress values and split the events into two equal-size groups. We find that for four out of the five bins, the \textit{b} value of the lower tidal stress group is larger than the higher tidal stress group. However, the statistical significance of the \textit{b} value differences is not guaranteed by the Utsu's \citep{Ut92} and \textit{z}-test due to the small number of events in each bin. 

\section{Discussions}
\label{sec:disc}
\citet{Sch15} calibrated the stress dependence of earthquake \textit{b} values assuming a simple frictional strength model combined with measurements of \textit{b} value variation with depth at different tectonic environments \citep{Spa13} and found that \textit{b} value varies by $\sim$0.001 MPa$^{-1}$. Our analysis suggests that earthquake \textit{b} value at Axial Seamount varies by $\sim$0.03 kPa$^{-1}$ (Fig. 3). In a recent tidal triggering study at Axial Seamount, \citet{Sch18} modeled the average Coulomb stress change on the 67$^{\circ}$ outward-dipping normal faults \citep{Lev18} (Fig. 1) due to vertical tidal stress changes as $\Delta CFS = \chi \sigma_{zz}$, with $\chi$ = 0.32 for a realistic magma chamber bulk modulus of 1 GPa. Adopting $\chi$ = 0.32 would give us a \textit{b} value change of $\sim$0.09 kPa$^{-1}$ of Coulomb stress change. Our observation of \textit{b} value variation that is sensitive to small stress perturbations ($\sim$10$^5$ more sensitive compared to \citet{Sch15}) is consistent with the long-documented observations of strong tidal triggering of earthquakes at Axial Seamount \citep{Tol02, Wil16} and other mid-ocean ridges \citep{Wil01, Str07}. \cite{Sch18} demonstrated that the seismicity rate change with tidal stress at Axial Seamount agrees with predictions of both rate-state and stress corrosion theories, and that the higher sensitivity can be explained by the shallow depths of the earthquakes (and hence the corresponding lower normal stress and stress drop values). The same explanation could apply to our observations since laboratory experiments previously showed that \textit{b} value variations depend on stress normalized to the maximum failure strength \citep{Sch68}. Our observed greater sensitivity is also consistent with observation of the \textit{b} value  of acoustic emissions in the laboratory varying with tidal stress \citep{Iwa05}.

While our calibrated \textit{b} value change with stress cannot be directly applied to other tectonic environments as most catalogued earthquakes occur at deeper depths, a sensitivity that is greater than 0.001 MPa$^{-1}$ \citep{Sch15} could explain observations of \textit{b} value decreasing preceding large earthquakes \citep{Nan12} and volcanic eruption \citep{Kat15}. Otherwise, these observations would represent stress changes on the order of 100 MPa in the decades before the Tohoku and Sumatra earthquakes \citep{Nan12} and weeks before the Mount Ontake eruption \citep{Kat15}. Alternatively, these documented \textit{b} value decreases might not have resulted from stress increases. Based on epidemic-type aftershock sequence (ETAS) modeling, \citet{Helm03} suggested that such \textit{b} value decreases can emerge simply from conditioning of the seismicity having to culminate in a mainshock, which results in there being a growing contribution of a deviatoric power law distribution with a smaller \textit{b} value to the background unconditional distribution. At Axial Seamount, \citet{Boh18} did not observe a systematic decrease in \textit{b} value leading up to the April 2015 eruption. \citet{Wil16} similarly did not observe the tidal triggering signal or the seismicity rate increasing leading up to the eruption. Therefore, the presumed stress accumulation in the three months before the volcano erupted might be too small to be detected with the current dataset.

Our observed stress dependence of earthquake \textit{b} values can be understood within the same statistical model first proposed to explain why the \textit{b} values of microfractures in laboratory experiments vary with stress \citep{Sch68}. If we treat the Earth's crust as an inhomogeneous elastic medium experiencing a uniform applied stress, the presence of inhomogeneities means that the stress at each point within the crust is a random variable that follows a probability distribution function that depends on the uniform applied stress. If we further assume that at each point, fracture will occur if the local stress exceeds a critical value and that fractures stop growing when they propagate into a region of lower stress, it follows that a fracture has a higher probability of growing larger when the applied stress is greater \citep{Sch68}. This translates to a decrease in \textit{b} value with increasing stress. However, the threshold effect that we observe is not well-explained by this model. Nevertheless, a similar threshold effect is often discussed for earthquake triggering from stress changes \citep{Roy96, Har98} with the stress threshold being dependent on the fault stiffness \citep{Roy96}.

A recent study using global data hinted at a \textit{b} value-tidal stress correlation, as earthquake \textit{b} values were found to decrease with increasing tidal shear stress ranking, where an earthquake's ranking is based on the maximum tidal shear stress during the day before the earthquake relative to the daily maxima in the 15 days before the earthquake \citep{Ide16}. However, the relationship was not clear for earthquakes smaller than $M_w$ 6.5 when looking at the Global Centroid Moment Tensor catalogue, potentially due to aggregating events of various faulting styles in diverse tectonic regimes \citep{Ide16}. Due to the lack of a strong correlation between global seismicity rate and tidal stress changes \citep[e.g.][and references therein]{Em97}, the authors instead invoke enhanced slow slip during increased tidal stresses that subsequently triggers earthquakes and increases the probability of rupture growth \citep{Ide16}. However, at Axial Seamount, the sensitivity of the earthquakes to small tidal stress changes \citep{Tol02, Wil16} can be simply explained by the shallow depths of the earthquakes \citep{Sch18} without invoking the existence of slow slip. 

\section{Conclusions}
Our natural experiment in a unique seafloor laboratory, looking at the size distribution of earthquakes in a 25 km$^3$ block of crust that experiences periodic tidal loading, provides a robust validation of the stress dependence of the earthquake \textit{b} value. We find that above a certain threshold stress amplitude, the earthquake \textit{b} value decreases linearly with increasing tidal stress. The \textit{b} value varies by $\sim$0.09 per kPa change in Coulomb stress. This suggests that \textit{b} value changes can be used to estimate stress variations in the Earth's crust. 

%%% End of body of article

%%%%%%%%%%%%%%%%%%%%%%%%%%%%%%%%%%%%%%%%%%%%%%%%%%%%%%%%%%%%%%%%
%
%  ACKNOWLEDGMENTS

\acknowledgments
Y. J. T. thanks G\"{o}ran Ekstrom, David Marsan, Meredith Nettles, Christopher Scholz, and Spahr Webb for fruitful discussions. This work was supported by NSF under grant OCE-1536320. The earthquake catalog used in this study is from \citet{Wil16} and is archived in the Interdisciplinary Earth Data Alliance Marine Geoscience Data System (DOI: 10.1594/IEDA/323843). The authors thank Suguru Yabe and two anonymous reviewers for their construtive comments that improved the manuscript. %The code for the analysis is available through https://github.com/yenjoetan. 

%%  REFERENCE LIST AND TEXT CITATIONS
%
% Either type in your references using
%
% \begin{thebibliography}{}
% \bibitem{}
% Text
% \end{thebibliography}
%
% Or, to use BibTeX:
%
% Follow these steps
%
% 1. Type in \bibliography{<name of your .bib file>} 

\begin{thebibliography}{}
\bibitem[{\textit{Agnew}}(1997)]{Ag97} Agnew, D. (1997), NLOADF: A program for computing ocean-tide loading, \textit{J. Geophys. Res., 102,} 5109--5110.

\bibitem[{\textit{Aki}}(1965)]{Aki65} Aki, K. (1965), Maximum likelihood estimate of b in the formula logN= a - bM and its confidence limits, \textit{Bull. Earthq. Res. Inst. 43}, 237--239.

\bibitem[{\textit{Albino et al.}}(2010)]{Alb10} Albino, F., Pinel, V., and Sigmundsson, F. (2010), Influence of surface load variations on eruption likelihood: application to two Icelandic subglacial volcanoes, Grimsvotn and Katla, \textit{Geophys. J. Int., 181}, 1510--1524.

\bibitem[{\textit{Amitrano}}(2003)]{Amit03} Amitrano, D. (2003), Brittle-ductile transition and associated seismicity: Experimental and numerical studies and relationship with the \textit{b} value, \textit{J. Geophys. Res., 108(B1)}, doi:10.1029/2001JB000680.

\bibitem[{\textit{Bohnenstiehl et al.}}(2008)]{Boh08} Bohnenstiehl, D. R., Waldhauser F. and Tolstoy, M. (2008), Frequency-magnitude distribution of microearthquakes beneath the 9\textdegree 50'N region of the East Pacific Rise, October 2003 through April 2004, \textit{Geochem. Geophys. Geosyst., 9(10)}, Q10T03, doi:10.1029/2008GC002128.

\bibitem[{\textit{Bohnenstiehl et al.}}(2018)]{Boh18} Bohnenstiehl, D. R. et al. (2018), Spatial, temporal and size-frequency characteristics of microearthquake sequences leading up to the 2015 eruption of Axial Seamount, \textit{Eos V52B-02}.

\bibitem[{\textit{Cao and Gao}}(2002)]{Cao02} Cao, A. M. and Gao S. S. (2002), Temporal variations of seismic b-values beneath northeastern japan island arc, \textit{Geophys. Res. Lett., 29}, doi:10.1029/2001GL013775.

\bibitem[{\textit{Cochran et al.}}(2004)]{Coch04} Cochran, E. S., Vidale, J. E., and Tanaka, S. (2004), Earth tides can trigger shallow thrust fault earthquakes, \textit{Science, 306}, 1164--1166, doi:10.1126/science.1103961.

\bibitem[{\textit{Egbert and Erofeeva}}(2002)]{Eg02} Egbert, G. D. and Erofeeva, S. Y. (2002), Efficient inverse modeling of barotropic ocean tides, \textit{J. Atmos. Ocean. Tech., 19}, 183--204.

\bibitem[{\textit{Emter}}(1997)]{Em97} Emter, D. (1997), Tidal triggering of earthquakes and volcanic events, in \textit{Tidal Phenomena}, edited by S. Bhattacharji et al., pp. 293--309, Springer, New York.

\bibitem[{\textit{Goebel et al.}}(2013)]{Goe13} Goebel, T. H. W., Schorlemmer, D., Becker, T. W., Dresen, G., and Sammis, C. G. (2013), Acoustic emissions document stress changes over many seismic cycles in stick-slip experiments, \textit{Geophys. Res. Lett., 40}, 2049--2054, doi:10.1002/grl.50507.

\bibitem[{\textit{Goebel et al.}}(2017)]{Goe17} Goebel, T. H. W., Kwiatek, G., Becker, T. W., Brodsky, E. E., and Dresen, G. (2017), What allows seismic events to grow big?: Insights from b-value and fault roughness analysis in laboratory stick-slip experiments, \textit{Geology, 45}, 815--818.

\bibitem[{\textit{Gulia et al.}}(2016)]{Gul16} Gulia, L., Tormann, T., Wiemer, S., Herrmann, M., and Seif, S. (2016), Short‐term probabilistic earthquake risk assessment considering time‐dependent b values, \textit{Geophys. Res. Lett., 43}, 1,100--1,108.

\bibitem[{\textit{Gutenberg and Richter}}(1944)]{Gut44} Gutenberg, B. and Richter, C. (1944), Frequency of earthquake in California, \textit{Bull. Seismol. Soc. Am., 34}, 185--188.

%\bibitem[{\textit{Hainzl}}(2016)]{Hai16} Hainzl, S. (2016), Rate-dependent incompleteness of earthquake catalogs, \textit{Seismol. Res. Lett., 87}, 337--344, doi:10.1785/0220150211.

\bibitem[{\textit{Hardebeck et al.}}(2003)]{Har98} Hardebeck, J. L., Nazareth, J. J., and Hauksson, E. (2003), The static stress change triggering model: Constraints from two southern California aftershock sequences, \textit{J. Geophys. Res., 103}, 24,427--24,437.

\bibitem[{\textit{Helmstetter et al.}}(2003)]{Helm03} Helmstetter, A., Sornette, D., and Grasso J. (2005), Mainshocks are aftershocks of conditional foreshocks: How do foreshock statistical properties emerge from aftershock laws, \textit{J. Geophys. Res., 108}, doi:10.1029/2002JB001991.

\bibitem[{\textit{Ide et al.}}(2016)]{Ide16} Ide S., Yabe, S., and Tanaka, Y. (2016), Earthquake potential revealed by tidal influence on earthquake size--frequency statistics, \textit{Nat. Geosci., 9}, 834--837, doi:10.1038/NGEO2796.

\bibitem[{\textit{Iwata and Young}}(2005)]{Iwa05} Iwata, T. and Young, R. P. (2005), Tidal stress/strain and the b-values of acoustic emissions at the Underground Research Laboratory, Canada, \textit{Pure appl. geophys, 162}, 1291--1308, doi: 10.1007/s00024-005-2670-2.

\bibitem[{\textit{Kagan}}(1997)]{Kag97} Kagan, Y. Y.  (1997), Seismic moment-frequency relation for shallow earthquakes: Regional comparison, \textit{J. Geophys. Res., 102}, 2835--2852.

\bibitem[{\textit{Kato et al.}}(2015)]{Kat15} Kato, A. et al. (2015), Preparatory and precursory processes leading up to the 2014 phreatic eruption of Mount Ontake, Japan, \textit{Earth, Planets and Space, 67}, https://doi.org/10.1186/s40623-015-0288-x.

\bibitem[{\textit{Levy et al.}}(2018)]{Lev18} Levy, S. et al. (2018), Mechanics of fault reactivation before, during, and after the 2015 eruption of Axial Seamount, \textit{Geology}, doi: https://doi.org/10.1130/G39978.1.

\bibitem[{\textit{Mignan and Woessner}}(2012)]{Mig12} Mignan, A. and Woessner, J. (2012), Estimating the magnitude of completeness for
earthquake catalogs, \textit{Community Online Resource for Statistical Seismicity Analysis}, doi:10.5078/corssa-00180805.

\bibitem[{\textit{Mori and Abercrombie}}(1997)]{Mor97} Mori, J. and Abercrombie, R. E. (1997), Depth dependence of earthquake frequency-magnitude distributions in California: Implications for rupture initiation, \textit{J. Geophys. Res., 102}, 15,081--15,090.

\bibitem[{\textit{Munk and Cartwright}}(1966)]{Munk66} Munk, W. H. and Cartwright, D. E. (1966), Tidal spectroscopy and prediction, \textit{Phil. Trans. Roy. Soc. Ser. A., 259}, 533--581.

\bibitem[{\textit{Nanjo et al.}}(2012)]{Nan12} Nanjo, K. Z., Hirata, N., Obara, K., and Kasahara, K. (2012), Decade-scale decrease in b value prior to the M9-class 2011 Tohoku and 2004 Sumatra quakes, \textit{Geophys. Res. Lett., 39}, L20304.

\bibitem[{\textit{Nishikawa and Ide}}(2014)]{Nish14} Nishikawa, T. and Ide, S. (2014), Earthquake size distribution in subduction zones linked to slab bouyancy, \textit{Nat. Geosci., 7}, 904--908, doi:10.1038/NGEO2279.

\bibitem[{\textit{Roy and Marone}}(1996)]{Roy96} Roy, M. and Marone, C. (1996), Earthquake nucleation on model faults with rate- and state-dependent friction: Effects of inertia, \textit{J. Geophys. Res., 101}, 13,919--13,923.

\bibitem[{\textit{Savcenko and Bosch}}(2012)]{Sav12}Savcenko, R. and Bosch, W.  (2012), EOT11a--Empirical ocean tide model from multi-mission satellite altimetry, \textit{DGFI Report No. 89}.

\bibitem[{\textit{Scholz}}(1968)]{Sch68} Scholz, C. H. (1968), The frequency-magnitude relation of microfracturing in rock and its relation to earthquakes, \textit{Bull. Seismol. Soc. Am., 58}, 399--415.

\bibitem[{\textit{Scholz}}(2015)]{Sch15} Scholz, C. H. (2015), On the stress dependence of the earthquake \textit{b} value, \textit{Geophys. Res. Lett., 42}, 1399--1402, doi:10.1002/2014GL062863.

\bibitem[{\textit{Scholz et al.}}(2018)]{Sch18} Scholz, C. H., Tan, Y. J., and Albino, F. (2018), The mechanism of tidal triggering of earthquakes at mid-ocean ridges, arXiv:1812.00639v1.

\bibitem[{\textit{Schorlemmer et al.}}(2005)]{Sch05a} Schorlemmer, D., Wiemer, S., and Wyss, M. (2005), Variations in earthquake-size distribution across different stress regimes, \textit{Nature, 437}, 539--542.

\bibitem[{\textit{Schorlemmer and Wiemer}}(2005)]{Sch05b} Schorlemmer, D. and Wiemer, S. (2005), Microseismicity data forecast rupture area, \textit{Nature, 434}, 1,086.

\bibitem[{\textit{Shi and Bolt}}(1982)]{Shi82} Shi, Y. and Bolt, B. A. (1982), The standard error of the magnitude-frequency b value, \textit{Bull. Seismol. Soc. Am., 72(5)}, 1677--1687.

\bibitem[{\textit{Spada et al.}}(2013)]{Spa13} Spada, M., Tormann, T., Wiemer, S., and Enescu, B. (2013), Generic dependence of the frequency-size distribution of earthquakes on depth and its relation to the strength profile of the crust, \textit{Geophys. Res. Lett., 40}, 709--714, doi:10.1029/2012GL054198.

\bibitem[{\textit{Stroup et al.}}(2007)]{Str07} Stroup, D. F., Bohnenstiehl, D. R., Tolstoy, M., Waldhauser, F., and Weekly, R. T. (2007), Pulse of the seafloor: Tidal triggering of microearthquakes at 9\textdegree 50'N East Pacific Rise, \textit{Geophys. Res. Lett., 34}, L15301, doi:10.1029/2007GL030088.

\bibitem[{\textit{Stumpf and Porter}}(2012)]{Stu12} Stumpf, M. and Porter, M. (2012), Critical truths about power laws, \textit{Science, 335}, 665--666.

\bibitem[{\textit{Tan et al.}}(2018)]{Tan18} Tan, Y. J., Tolstoy, M., Waldhauser, F.,  and Bohnenstiehl, D. R. (2018), Tidal triggering of microearthquakes over an eruption cycle at 9\textdegree 50'N East Pacific Rise, \textit{Geophys. Res. Lett., 45}, 1825--1831.

\bibitem[{\textit{Tolstoy et al.}}(2002)]{Tol02} Tolstoy, M., Vernon, F. L., Orcutt, J. A., and Wyatt, F. K. (2002), Breathing of the seafloor: Tidal correlations of seismicity at Axial Volcano, \textit{Geology, 30}, 503--506.

\bibitem[{\textit{Tr{\'e}hu and Solomon}}(1983)]{Tre83} Tr{\'e}hu, A. M. and Solomon, S. C. (1983), Earthquakes in the Orozco Transform Zone: Seismicity, source mechanisms, and tectonics, \textit{J. Geophys. Res., 88(B10),} 8203--8225.

\bibitem[{\textit{Utsu}}(1966)]{Ut66} Utsu, T. (1966), A statistical significance test of the difference in b-value between two earthquake groups, \textit{J. Phys. Earth, 14}, 34--40.

\bibitem[{\textit{Utsu}}(1992)]{Ut92} Utsu, T. (1992), On seismicity, \textit{Rep. Jt. Res. Inst. Stat. Math., 34}, 139--157.

%\bibitem[{\textit{Utsu}}(1999)]{Ut99} Utsu, T. (1999), Representation and analysis of the earthquake size distribution: A historical review and some new approaches, \textit{Pure Appl. Geophys., 155}, 509--535.

\bibitem[{\textit{Vidale et al.}}(1998)]{Vid98} Vidale, J., Agnew, D., Johnston, M., and Oppenheimer, D. (1998), Absence of earthquake correlation with earth tides: An indication of high preseismic fault stress rate, \textit{J. Geophys. Res., 103(24),} 567--572.

\bibitem[{\textit{Waldhauser and Ellsworth}}(2000)]{Wald00} Waldhauser, F. and Ellsworth, W. L. (2000), A double-difference earthquake location and algorithm: Method and application to the northern Hayward Fault, California, \textit{Bull. Seismol. Soc. Am., 90}, 1353--1368, doi:10.1785/0120000006.

%\bibitem[{\textit{Waldhauser and Tolstoy}}(2011)]{Wald11} Waldhauser, F., and M. Tolstoy (2011), Seismogenic structure and processes associated with magma inflation and hydrothermal circulation neneath the East Pacific Rise at 9\textdegree 50'N, \textit{Geochem. Geophys. Geosyst., 12(9)}, Q08T10, doi:10.1029/2011GC003568.

\bibitem[{\textit{Wang and Shearer}}(2015)]{Wang15} Wang, W. and Shearer, P. M. (2015), No clear evidence for localized tidal periodicities in earthquakes in the central Japan region, \textit{J. Geophys. Res., 120,} 6317--6328, doi:10.1002/2015jb011937.

\bibitem[{\textit{Wiemer and Katsumata}}(1999)]{Wie99} Wiemer, S. and Katsumata, K. (1999), Spatial variability of seismicity parameters in aftershock zones, \textit{J. Geophys. Res., 103,} 13135--13151.

\bibitem[{\textit{Wiemer and Wyss}}(2000)]{Wie00} Wiemer, S. and Wyss, M. (2000), Minimum magnitude of completeness in earthquake catalogs: Examples from Alaska, the western United States, and Japan, \textit{Bull. Seismol. Soc. Am., 90}, 859--869, doi:10.1785/0119990114.

%\bibitem[{\textit{Wiemer and Wyss}}(2002)]{Wie02} Wiemer, S., and M. Wyss (2002), Mapping spatial variability of the frequency-magnitude distribution of earthquakes, \textit{Adv. Geophys., 45}, 259--302.

\bibitem[{\textit{Wilcock}}(2001)]{Wil01} Wilcock, W. S. D. (2001), Tidal triggering of microearthquakes on the Juan de Fuca Ridge, \textit{Geophys. Res. Lett., 28}, 3999--4002.

\bibitem[{\textit{Wilcock et al.}}(2016)]{Wil16} Wilcock, W. S. D. et al. (2016), Seismic constraints on caldera dynamics from the 2015 Axial Seamount eruption, \textit{Science, 354(6318)}, 1395--1399.

\bibitem[{\textit{Wilcock et al.}}(2018)]{Wil18} Wilcock, W. S. D. et al. (2018), The recent volcanic history of Axial Seamount: Geophysical insights into past eruption dynamics with an eye toward enhanced observations of future eruptions, \textit{Oceanography, 31(1)}, 114--123.

\bibitem[{\textit{Woessner and Wiemer}}(2005)]{Woes05} Woessner, J. and Wiemer, S. (2005), Assessing the quality of earthquake catalogues: Estimating the magnitude of completeness and its uncertainty, \textit{Bull. Seismol. Soc. Am., 95}, 684--698, doi: 10.1785/0120040007.

\end{thebibliography}
%    Run LaTeX on your LaTeX file.
%
% 2. Run BiBTeX on your LaTeX file.
%
% 3. Open the new .bbl file containing the reference list and
%   copy all the contents into your LaTeX file here.
%
% 4. Run LaTeX on your new file which will produce the citations.
%
% AGU does not want a .bib or a .bbl file. Please copy in the contents of your .bbl file here.

%%%%%%%%%%%%%%%%%%%%%%%%%%%%%%%%%%%%%%%%%
% Track Changes:
% To add words, \added{<word added>}
% To delete words, \deleted{<word deleted>}
% To replace words, \replace{<word to be replaced>}{<replacement word>}

% At the end of the document, use \listofchanges, which will list the
% changes and the page and line number where the change was made.

% When final version, \listofchanges will not produce anything,
% \added{} word will be printed, \deleted{} will take away the word,
% \replaced{}{} will print only the 2nd argument.

%%%
%%%

\end{document}

%% This command needs article title as argument to \supportinginfo{}:
\supportinginfo{Axial Seamount: Periodic tidal loading reveals stress dependence of the earthquake size distribution (\textit{b} value)}

 \authors{Yen Joe Tan\affil{1},
 Felix Waldhauser\affil{1}, Maya Tolstoy\affil{1}, and William S. D. Wilcock\affil{2}}
 
\affiliation{1}{Lamont-Doherty Earth Observatory of Columbia University, Palisades, New York 10964, USA.}
\affiliation{2}{School of Oceanography, University of Washington, Seattle, WA
98195, USA.}
%% Corresponding Author
%(include name and email addresses of the corresponding author.  More
%than one corresponding author is allowed in this Word file and for
%publication; but only one corresponding author is allowed in our
%editorial system.)  
\correspondingauthor{Yen Joe Tan}{yjt@ldeo.columbia.edu}
%\section*{Contents}
\begin{enumerate}
\item Figure S1

Selection of the minimum magnitude of completeness ($M_c$). \textbf{top}, Cumulative (triangle) and non-cumulative (circle) frequency-magnitude distributions (FMDs). Dashed lines depict the $M_c$ estimates based on the MAXC (green), GFT-90\% (blue), GFT-95\% (red), and MBS (magenta) methods. \textbf{middle}, Variation of parameter \textit{R} used to quantify the goodness-of-fit between observed and synthetic cumulative FMDs for a range of cutoff magnitudes. Dashed lines depict the 90\% (blue) and 95\% (red) thresholds. \textbf{bottom}, Variation of the \textit{b} value for a range of cutoff magnitudes. Dashed line depicts where the \textit{b} value first stabilizes (see Methods). 

\centerline{\includegraphics[height=6in]{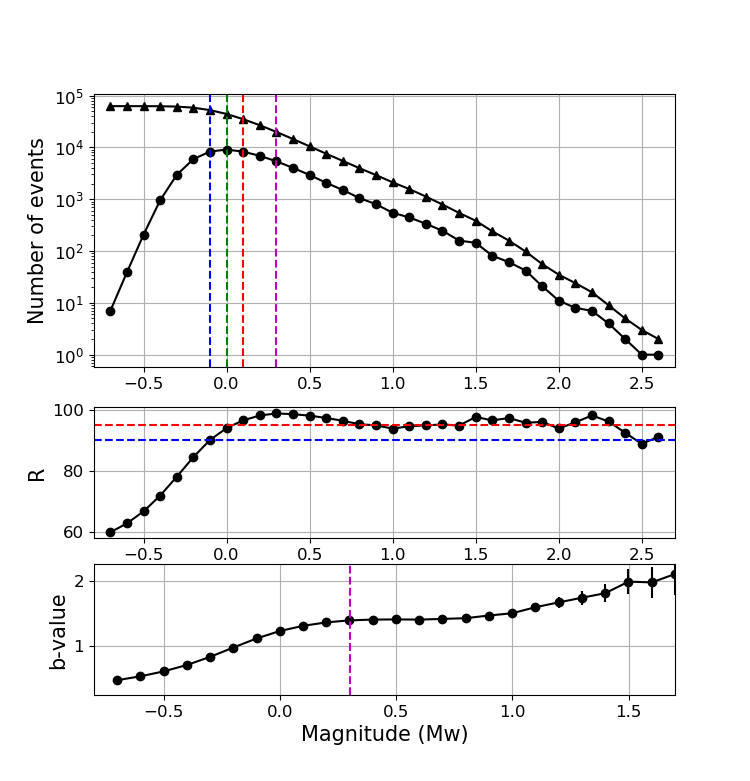}}

\item Figure S2

Time series of predicted tides at the surface at 45.95$^\circ$N, 130.00$^\circ$W. \textbf{top}, Estimate of $\sigma_{xx}$ (black), $\sigma_{yy}$ (blue), and $\sigma_{zz}$ (red) from ocean tidal loading, with tension being positive (i.e. $\sigma_{zz}$ is positive upwards). \textbf{bottom}, Body tides. The stress amplitude is about an order of magnitude smaller than that of ocean tides. 

\centerline{\includegraphics[height=4.5in]{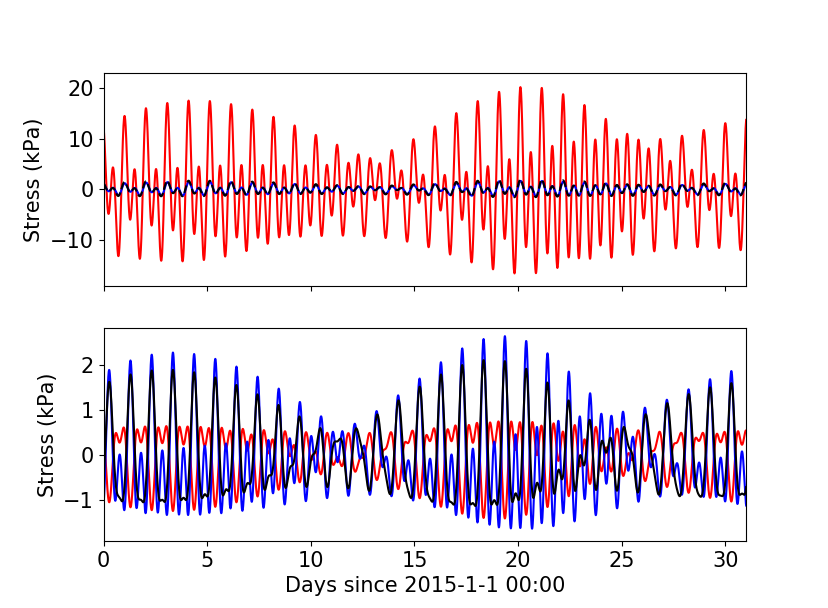}}

\item Figure S3

The earthquake tidal stress distribution in bins of 2 kPa for $M_c$ = 0.1 (blue) and $M_c$ = 0.3 (grey). The distribution reflects the combined effect of seismicity rate increasing with tidal stress \citep{Sch18} and the uneven distribution of tidal stress amplitudes. 

\centerline{\includegraphics[height=4.5in]{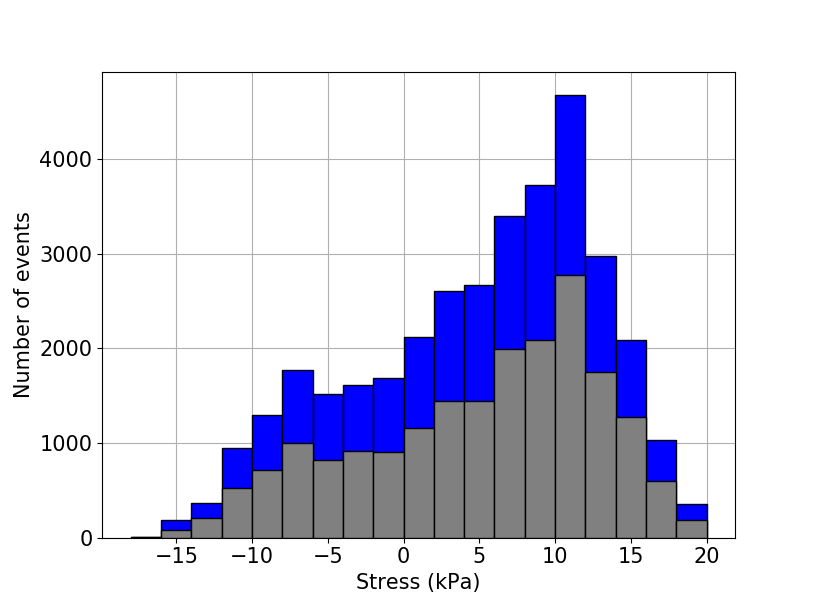}}

\item Figure S4

The earthquake \textit{b} values for non-overlapping stress bins of 2 kPa. Earthquakes of $M_w$ greater than 1.5 are excluded. The vertical error bars represent two standard deviations of the estimated \textit{b} values from bootstrapping. The horizontal bars represent the tidal stress range for each bin. Dashed lines represent linear least-squares fits, both giving \text{b} value varying by $\sim$ 0.03 per kPa. \textbf{a}, Using $M_c$ = 0.1. \textbf{b}, Using $M_c$ = 0.3.

\centerline{\includegraphics[height=3.5in]{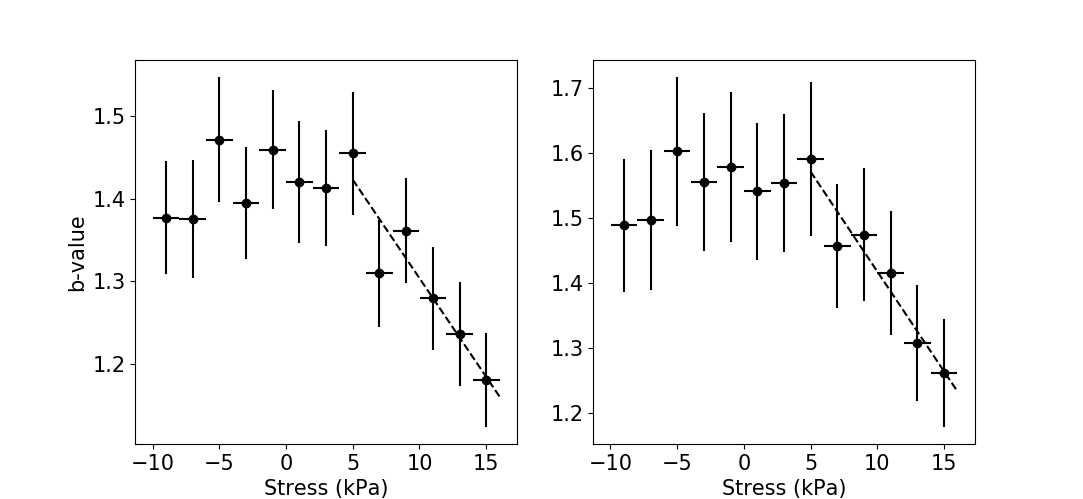}}

\item Figure S5

The earthquake \textit{b} value as a function of depth in non-overlapping bins of 2,500 earthquakes. The vertical error bars represent two standard deviations \citep{Shi82} of the estimated \textit{b} values. The horizontal bars represent the range of earthquake depth values included in each bin. Since $M_c$ is expected to vary with depth, we estimate $M_c$ for different depth ranges using the GFT-95\% method before choosing a fixed $M_c$ = 0.3 to estimate the \textit{b} values.

\centerline{\includegraphics[height=3.5in]{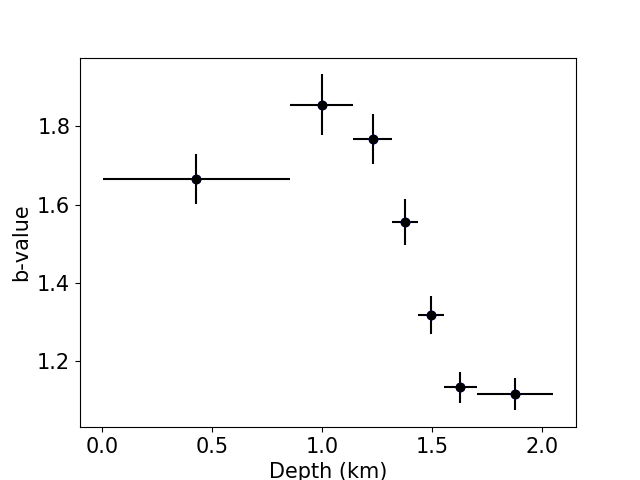}}

\item Figure S6

\textbf{a}, Cumulative number of events with depth for the lower tidal stress group (blue) and higher tidal stress group (red) (see Fig. 4). \textbf{b}, Mean (circle) and median (triangle) earthquake depth for non-overlapping stress bins of 4 kPa.

\centerline{\includegraphics[height=3.5in]{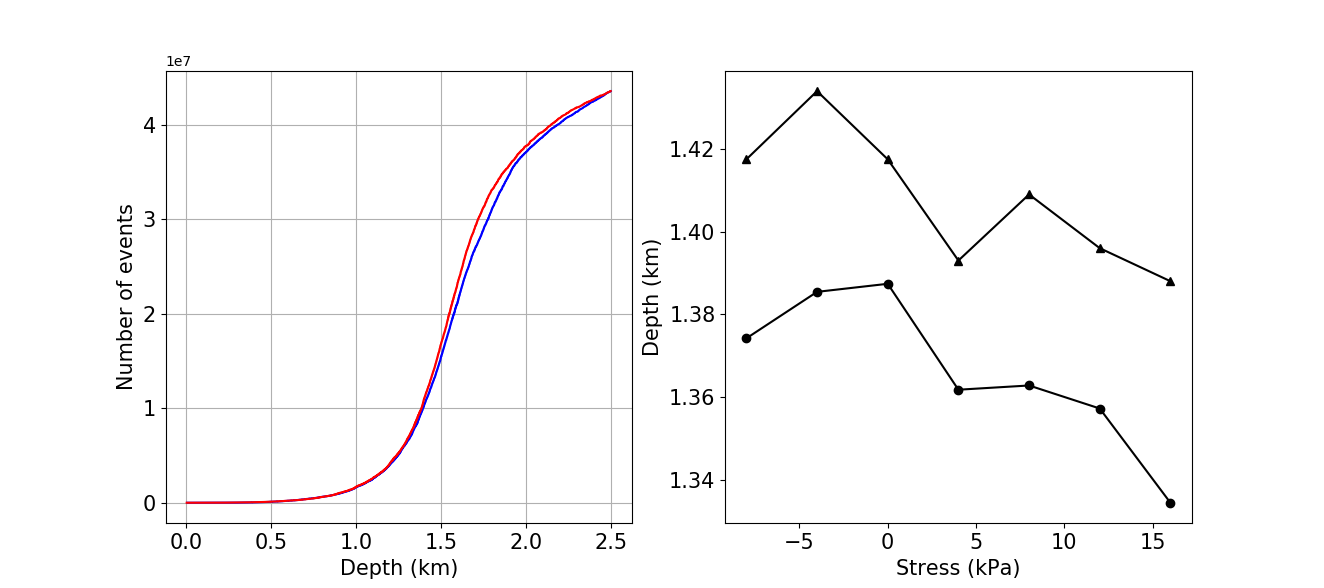}}

\item Figure S7

The normalized earthquake spatial distribution in 1 km$^2$ grids for \textbf{a}, lower tidal stress group and \textbf{b}, higher tidal stress group (see Fig. 4). \textbf{c}, The earthquake spatial distribution in 1 km$^2$ grids for all events above $M_c$ = 0.3. Only grids containing more than 1,000 events are shown. 

\centerline{\includegraphics[height=4.5in]{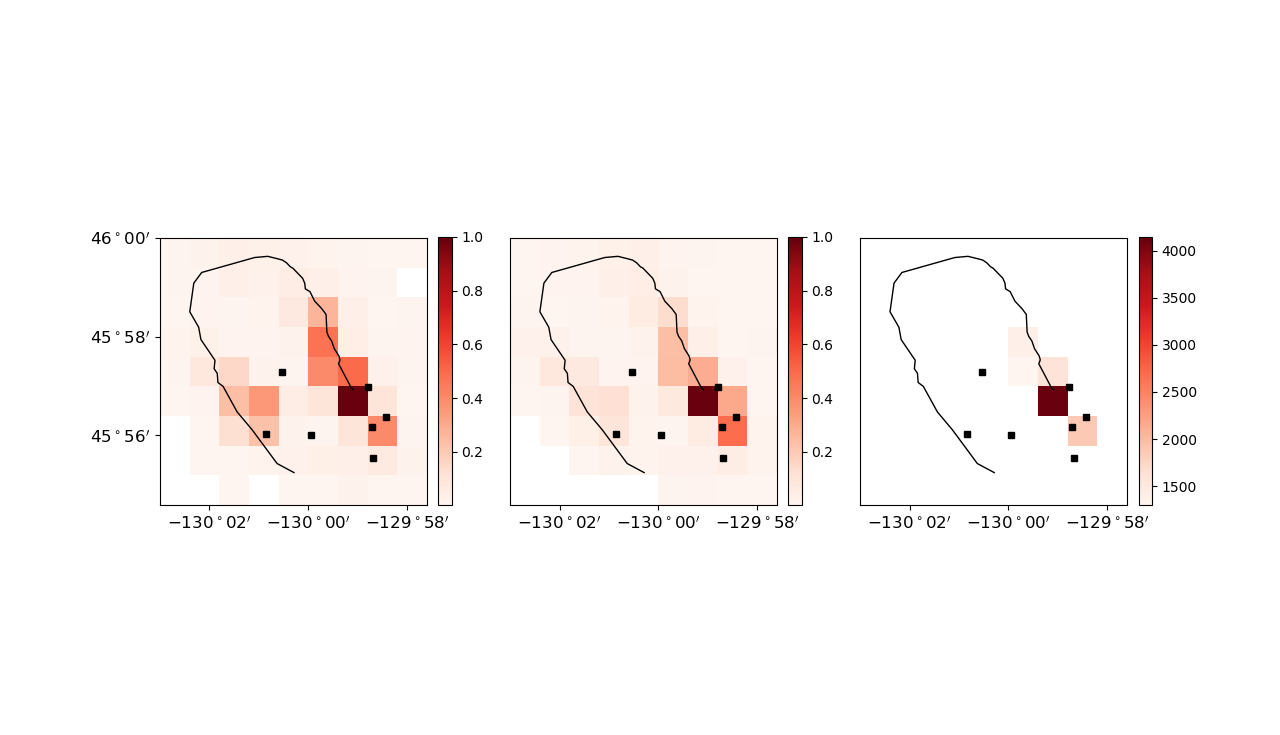}}

\end{enumerate}

{}